\pdfoutput=1

\documentclass[11pt,twoside,a4paper,cmspaper,final,collab]{cms-tdr}
\def\svnVersion{167918}\def\cmsCernNo{2012-166}\def\cmsCernDate{\today}\def\cmsMessage{Submitted to Physics Letters B}

\begin{document}\cmsNoteHeader{EXO-11-056}
\typeout{**** Begin including local definitons.tex file *****}
\typeout{**** tdr loads this quietly without a specific input command *****}

\newcommand{\fix}[1]{{\bf <<< #1 !!!} }
 
\typeout{**** End including local definitons.tex file *****}

\def\roostats{\texttt{RooStats}\xspace}
\def\L{{\ensuremath{\cal L}}\xspace}
\def\bnu    {\ensuremath{{\boldsymbol\nu}}\xspace}
\def\bsigma {\ensuremath{{\boldsymbol\sigma}}\xspace}
\def\selEff{\ensuremath{\epsilon_\mathtt{sel}}\xspace}
\def\acc{\ensuremath{{\cal A}}\xspace}

\def\proton      {\ensuremath{p}\xspace}

\hyphenation{had-ron-i-za-tion}
\hyphenation{cal-or-i-me-ter}
\hyphenation{de-vices}

\RCS$Revision: 144342 $
\RCS$HeadURL: svn+ssh://svn.cern.ch/reps/tdr2/papers/EXO-11-056/trunk/EXO-11-056.tex $
\RCS$Id: EXO-11-056.tex 144342 2012-08-18 09:53:57Z cvuosalo $
\newlength\cmsFigWidthA
\newlength\cmsFigWidthB
\ifthenelse{\boolean{cms@external}}{\setlength\cmsFigWidthA{0.45\columnwidth}}{\setlength\cmsFigWidthA{0.3\textwidth}}
\ifthenelse{\boolean{cms@external}}{\setlength\cmsFigWidthB{\columnwidth}}{\setlength\cmsFigWidthB{0.67\textwidth}}
\ifthenelse{\boolean{cms@external}}{\providecommand{\cmsLeft}{top}}{\providecommand{\cmsLeft}{left}}
\ifthenelse{\boolean{cms@external}}{\providecommand{\cmsRight}{bottom}}{\providecommand{\cmsRight}{right}}
\newcommand{\pwprp}{\ensuremath{\mathrm{W^{'+}}}}
\newcommand{\pwprm}{\ensuremath{\mathrm{W^{'-}}}}
\cmsNoteHeader{EXO-11-056} 
\title{Search for charge-asymmetric production of \PWpr\ bosons in $\ttbar + \text{jet}$
events from $\Pp\Pp$ collisions at $\sqrt{s} = 7\TeV$}

\date{\today}

\abstract{A search is presented for
charge-asymmetric production of a \PWpr\ boson
that has been proposed to accommodate the forward-backward asymmetry
observed in the production of top-antitop quark pairs
at the Tevatron.
The new heavy \PWpr\ boson would be produced in association with
a top quark and would
decay into top and down quarks.
The data
correspond to an integrated luminosity of 5.0\fbinv in pp
collisions at a center-of-mass energy of 7\TeV,
recorded by the CMS detector at the LHC.
No significant excess above the standard model expectations is observed,
and, from a combination of the electron-plus-jets and muon-plus-jets channels,
a 95\% confidence level lower limit of 840\GeVcc is set on the \PWpr\ boson
mass for a \PWpr\ boson model with values for 
coupling constants to top and down quarks $g_{L}=0$ and $g_{R}=2$.
In addition, a
kinematic reconstruction of the $\PWpr$ resonance mass
using the inherent charge asymmetry of this model
finds no indication of the presence of $\PWpr$ events in the data.
}

\hypersetup{%
pdfauthor={CMS Collaboration},%
pdftitle={Search for charge-asymmetric production of W' bosons in top pair + jet
events from pp collisions at sqrt(s) = 7 TeV},%
pdfsubject={CMS},%
pdfkeywords={CMS, W', top pair}}

\maketitle 

\section{Introduction}
Many extensions to the standard model (SM) involve enhanced gauge symmetries
that give rise to additional gauge bosons.
Within these extensions, one of the additional bosons could be the $\PWpr$ boson,
a proposed heavy partner to the $\PW$ boson.
There are many scenarios with a $\PWpr$ boson:
a left-right symmetric model~\cite{wprimeleftright},
a model based upon a new SU(2) sector~\cite{wprimeleftright},
a technicolor model~\cite{wprimetechnicolor},
and a $\PWpr$ as the lowest Kaluza--Klein mode of the $\PW$ boson~\cite{wprimekkmode}.

\begin{figure}[htbp]
\begin{center}
\includegraphics[width=\cmsFigWidthA]{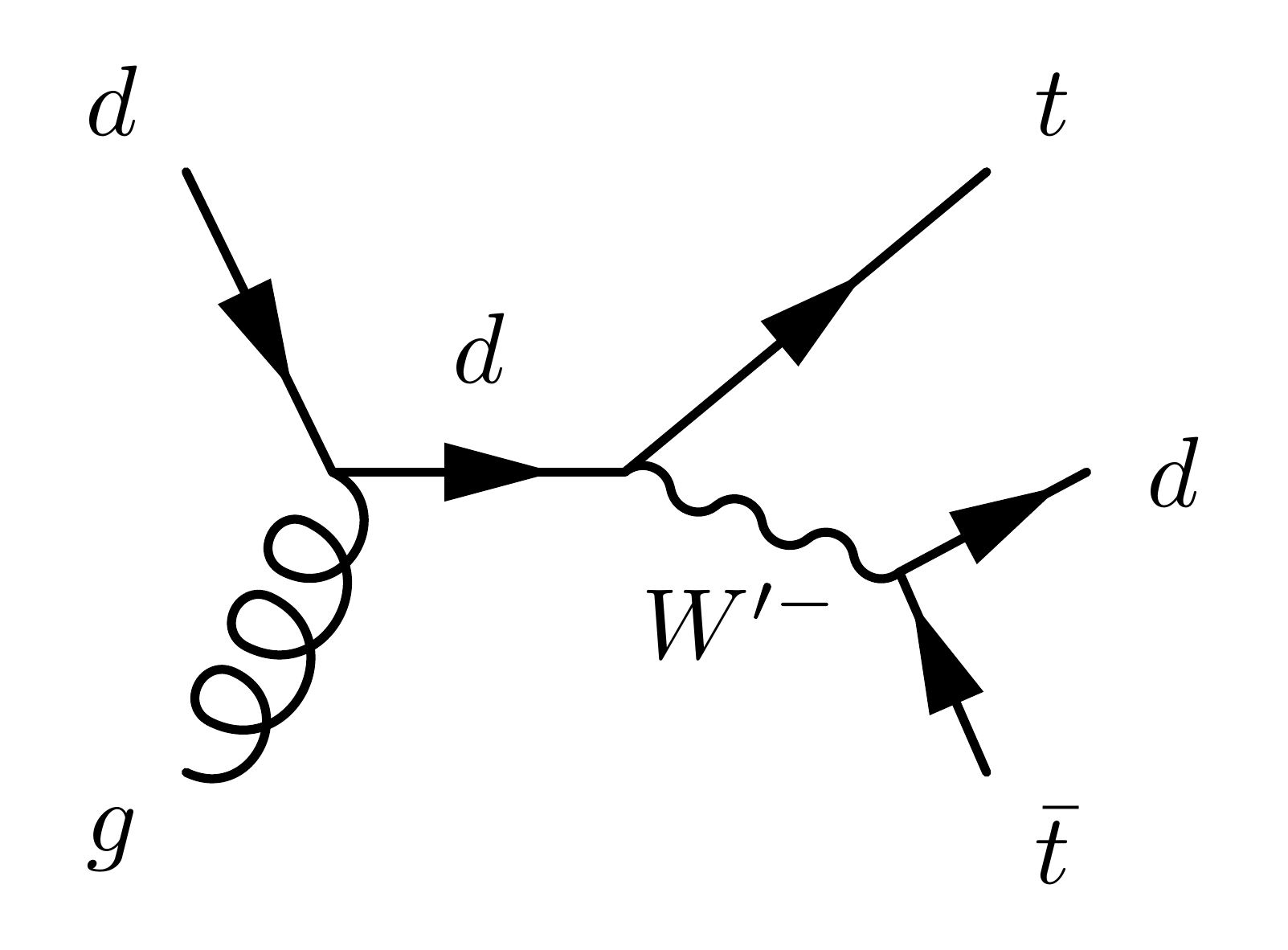}
\includegraphics[width=\cmsFigWidthA]{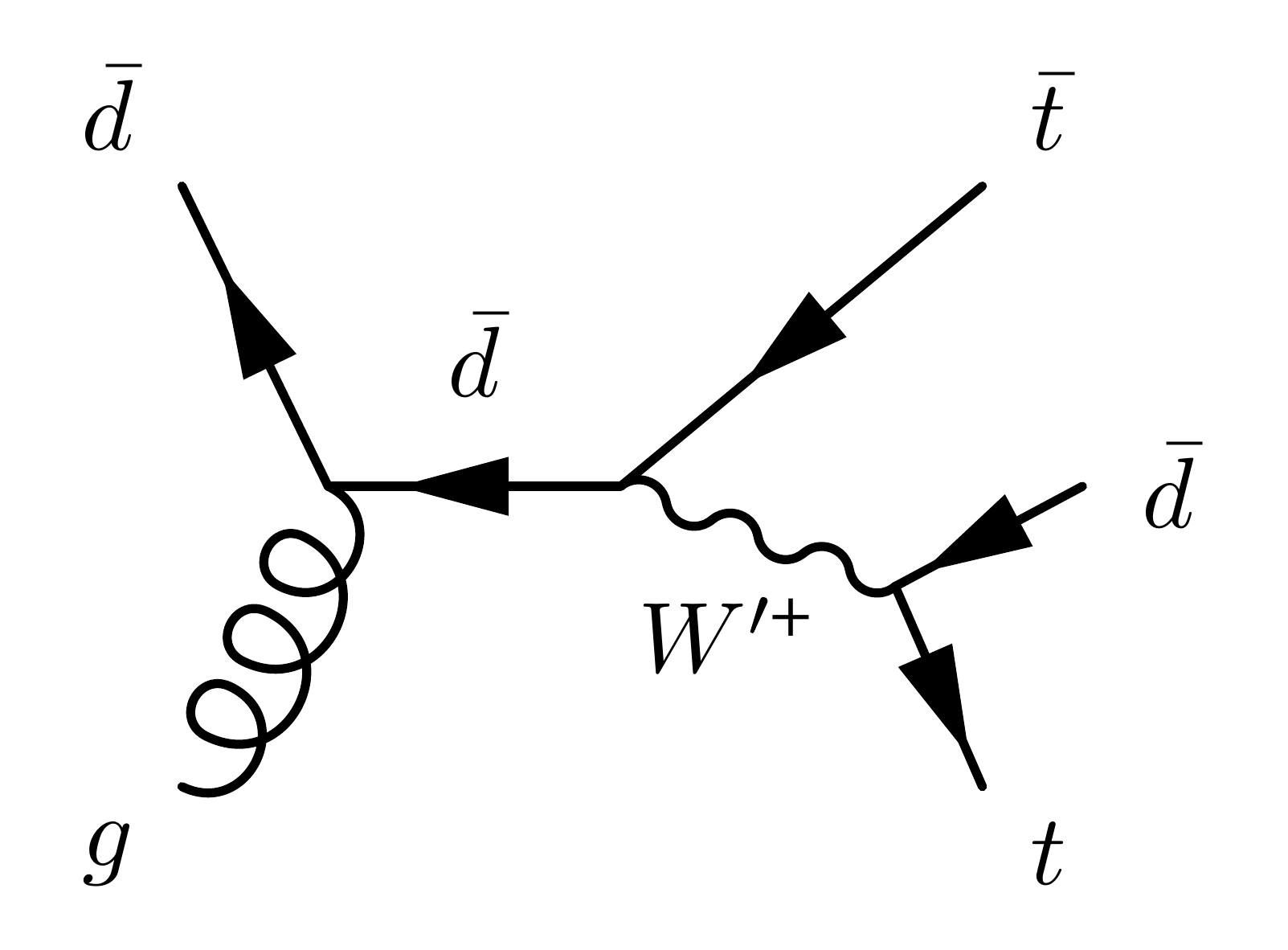}
\caption{Feynman diagrams for $s$-channel production of (left) $\pwprm$ and (right) $\pwprp$. Diagrams for $t$-channel production
can be found in Ref.~\cite{Strassler}.}
\label{fig:wpfeynman}
\end{center}
\end{figure}

A  $\PWpr$ boson with a coupling to top (t) and down (d) quarks has been proposed~\cite{wprime}
to explain the anomalous
forward-backward asymmetry in \ttbar
events reported at the
Tevatron~\cite{PhysRevLett.100.142002, PhysRevD.83.112003, PhysRevD.84.112005}.
The observed effect, which
is particularly significant for large values of the \ttbar mass,
could be explained by the production of a $\PWpr$ boson with
a mass in the range of 200--600\GeVcc~\cite{wprime}.
A search for a $\PWpr$
decaying to top and light quarks was conducted by the CDF experiment~\cite{Aaltonen:2012qn}.
With a predicted cross section around 20\unit{pb} at 7\TeV
and an assumed 100\% branching fraction into t and d quarks,
as illustrated in Fig. \ref{fig:wpfeynman}, the $\PWpr$ boson is potentially observable with the data already collected by the Compact Muon Solenoid (CMS) collaboration at the Large Hadron Collider (LHC).

Because the LHC collides protons with protons, rather than with antiprotons,
the valence \cPqd\ quarks
have a much larger fraction of beam particle momentum than
$\cPaqd$ quarks, which come from the proton ``sea''.  The result is
that, at leading order (LO),
the $\pwprm$ ($\pwprp$) contributes about 85\% (15\%) of the
total $\PWpr$ production cross section,
for $\PWpr$ masses in the range from 400--1200\GeVcc~\cite{Strassler}.
This feature can be used to aid in the identification of the $\PWpr$, as explained in Section \ref{Sec:CS_strategy}.

The LO processes shown
in Fig.~\ref{fig:wpfeynman} result in a final state of \ttbar plus a \cPqd\ quark or antiquark.
This final state
can be classified according to how the $\PW$ bosons from the top quarks decay:
all hadronic (both $\PW$ bosons decaying hadronically), partially leptonic (one
decaying hadronically, the other leptonically), or fully leptonic
(both decaying leptonically).
We focus on the partially leptonic mode because it has a larger branching fraction than the
fully leptonic mode and a cleaner signature than the
all-hadronic mode.
The event selection for this analysis requires one electron or muon accompanied by several
jets and an imbalance in transverse momentum.
The main background originates from SM \ttbar production with
initial- or final-state radiation.
We conduct a search for the $\PWpr$ signal by comparing the number of
observed events in data with the total expected from SM sources.
In addition, we utilize a
kinematic reconstruction of the $\PWpr$ resonance mass
and the inherent charge asymmetry of this model
to perform an independent test for the presence of $\PWpr$ events in the data.

\section{The CMS Detector and Data Samples}
The central feature of the CMS apparatus is a superconducting solenoid, of 6\unit{m} internal diameter, providing a magnetic field of 3.8\unit{T}. A silicon pixel and strip tracker resides within the
field volume, surrounded by a lead tungstate crystal electromagnetic calorimeter (ECAL)
and a brass/scintillator hadron calorimeter. Muons are
measured in gas-ionization detectors embedded
in the flux-return yoke of the solenoid. Extensive forward calorimetry complements the coverage provided by the barrel and endcap
detectors.
CMS uses a right-handed coordinate system, with the origin at the nominal interaction point, the $x$ axis pointing to the center of the
LHC ring, the $y$ axis pointing up (perpendicular to the plane of the LHC ring), and the $z$ axis along the counterclockwise beam direction.
The polar angle, $\theta$, is measured from the positive $z$ axis and the azimuthal angle,
$\phi$, is measured in the $x$-$y$ plane.
The pseudorapidity is defined as $\eta =  - \ln[\tan(\theta/2)]$.
A more detailed description of the detector can be found in Ref.~\cite{cms}.

The data sample corresponds to an integrated luminosity of
5.0$\pm 0.1$\fbinv in pp collisions at a center-of-mass energy $\sqrt{s}=7$\TeV, collected
by the CMS detector at the LHC.
We perform our search in both the electron-plus-jets (e+jets) and
muon-plus-jets ($\mu$+jets) channels.
For the e+jets channel, about 0.2\fbinv of data
were initially collected by requiring one electron with transverse
momentum ($\pt$) greater than 27\GeVc.
For the next 1\fbinv of data, the  $\pt$ threshold had to be raised to
32\GeVc.
During the course of data taking, the maximum instantaneous
luminosity increased by an order of magnitude, reaching up to $4\times 10^{33}\percms$,
thereby requiring changes in the trigger configurations.
For the remaining 3.8\fbinv of data, it was necessary to include
several jets in the electron trigger, resulting in
an electron $\pt$ threshold of 25\GeVc plus three jets
with $\pt>30\GeVc$.
For the $\mu$+jets channel, 2.2\fbinv of data
were initially collected by requiring a single muon with $\pt>30\GeVc$, with
this threshold later raised to 40\GeVc
for the last 2.8\fbinv of data.

The data are compared with simulations of SM background
contributions from leptonically enriched multijets, single top quark, \ttbar, W, and Z production,
with all sources including additional jets.
The \ttbar background is dominant, and is simulated with up to three
additional partons by
the \MADGRAPH 4.4.12~\cite{ref:madgraph} event generator interfaced with the
\PYTHIA 6.4.22~\cite{ref:pythia} parton shower simulator.
In the matching
procedure for this parton showering~\cite{mlm},
the $k_\mathrm{T}$ matrix element uses a matching scale of 30\GeVc, according to
the MLM scheme~\cite{Mangano}.
The $\PW$ and $\cPZ$ background processes are produced with the \MADGRAPH event generator.
The $\PWpr$ signal samples are produced with \MADGRAPH
for masses $M_{\PWpr}$ = 400, 600, 800, 900,
1000, and 1200\GeVcc with
values of the $\PWpr$ coupling constants to top and down quarks $g_{L}=0$ and $g_{R}=2$~\cite{Strassler}.
Additional $\PWpr$ benchmark points are produced
for masses $M_{\PWpr}$ = 600 and 800\GeVcc with $g_{L}=0$ and $g_{R} = \sqrt{2}$. Compared to Ref.~\cite{Aaltonen:2012qn}, the definition of $g_{R}$ used here gives values smaller by a factor of $1/\sqrt{2}$.
Single top quark production is simulated with the \POWHEG~\cite{bib:powheg}
event generator and includes s- and t-channel production, along with tW
associated production.
The multijet background contribution is simulated using a combination of two sets of samples, one generated
with \MADGRAPH and the other with \PYTHIA.
The CTEQ6L1 parton distribution function (PDF) set \cite{CTEQ6L1},
which are LO PDFs, is used for generating all simulated events.
The events for all samples are passed through a
\GEANTfour-based simulation~\cite{ref:geant4} of the CMS detector and
reconstructed with the same program used to reconstruct data.

\section{Event Selection and Reconstruction}
\label{sec:evtsel}
The particle-flow (PF) algorithm~\cite{bib:Pflow} is used to reconstruct and identify
each particle based upon an optimized combination of information from all the sub-detectors.
The particles are classified as charged hadrons, neutral hadrons, photons, muons, or electrons.

Charged leptons originating from $\PW$ boson decays are expected to be isolated from other particles in the event.
A variable to quantify this lepton isolation, $I$, is defined as a sum
of momenta, divided by the lepton $\pt$, where the sum is of the
transverse momenta of charged
hadrons, neutral hadrons, and photons in a cone of $\Delta R \equiv \sqrt{(\Delta \eta)^2+(\Delta \phi)^2} < 0.3$ around the
lepton direction, excluding the contribution from the lepton itself.
Muons are reconstructed using information from the silicon tracker and the muon chambers.
Muon candidates are required to be isolated, with $I<0.125$,
and to have $\pt > 42$\GeVc and  $|\eta| < 2.1$.
Electrons are reconstructed using associated clusters of energy deposits in the ECAL that are then combined
with tracks from the inner tracker \cite{CMS-PAS-EGM-10-004}.
The electron candidate is required to be isolated, with $I<0.1$, and to have  $\pt>45\GeVc$ and $| \eta | < 2.5$.
The $\pt$ thresholds for the leptons are
chosen so as to ensure a $\pt$-independent trigger efficiency,
whose value is about 98\%, for selected events.
Electrons in the $\eta$ range around the interface between the ECAL endcap and ECAL barrel ($1.444 < |\eta| < 1.566$) are excluded from the selection.
From the high-quality vertices that are
close to the beam spot, the vertex with the highest sum of $\pt^{2}$ of its constituent tracks
is chosen as the primary vertex.
To ensure the reconstructed muon or electron track is consistent
with originating from the primary vertex, we require the
longitudinal distance between the track and vertex
to be less than 1\unit{cm}, and the point of closest
approach to the primary vertex in the transverse direction to
be less than 0.02\unit{cm}, as suggested by Ref.~\cite{topref}.
Selected events must contain exactly one electron or muon meeting the above requirements, and
events are rejected if they have any additional muons with $\pt>10\GeVc$,
$|\eta|<2.5$, and $I<0.2$ or electrons with $\pt>20\GeVc$,
$|\eta|<2.5$, and $I<0.2$.
To remove electrons resulting from photon conversions,
we eliminate those events with a single electron candidate
that has no hits in the innermost pixel layer or that is accompanied
by a nearby track.

Jets are reconstructed from PF constituents
with the \textsc{FastJet} package~\cite{fastjet}, using the anti-$k_\mathrm{T}$~\cite{antikt}
clustering algorithm with a distance parameter 0.5 and a
jet clustering recombination scheme
that merges particles by summing their four-vectors~\cite{Cacciari:2011ma}.
Only charged PF jets that are consistent with originating from
the primary vertex are considered.
Correction factors~\cite{JES_official} are applied to the jet energy
to account for non-linearities in detector response.

Selected events must have at least five jets with a minimum $\pt$ of 35\GeVc and $|\eta| < 2.4$. The highest-$\pt$ jet is required to have $\pt > 180\GeVc$, and
the second-highest-$\pt$ jet must have  $\pt > 90\GeVc$. The requirement for
the highest-$\pt$ jet is intended, in part, to select for the d-quark jet coming
from the \PWpr\ decay.
We also require $H_\mathrm{T}>700\GeVc$, where $H_\mathrm{T}$
is the scalar sum of all jet $\pt$, the lepton $\pt$, and a
quantity called \ETm, which is the absolute value of the
vector sum of the transverse momenta
of all particles found by the PF algorithm, with the particles being treated
as massless.
These $H_\mathrm{T}$ and jet $\pt$values are determined from optimizing the selection to
suppress SM backgrounds while enhancing the signal significance,
which is taken to be the expected number of signal events for a benchmark point with a $\PWpr$
mass of 600\GeVcc divided by the square root of the number of expected background events.  This benchmark corresponds to
the highest mass point that is
able to account well for the forward-backward asymmetry in \ttbar events
observed at the Tevatron~\cite{wprime}.
We further require that at least one jet is identified as originating from a  b quark.
Jets from b quarks are identified by
an algorithm~\cite{CMS-PAS-BTV-11-001} that reconstructs a displaced secondary vertex
with high efficiency by combining two or more tracks and then assigns
a likelihood of b-quark origin based upon the three-dimensional decay
length of the vertex.

The leptonic decay of the $\PW$ boson arising from a top quark
produces a neutrino, which escapes the detector without interacting.
The \ETm provides a measure of this missing energy, so
we require $\ETm>20\GeV$.

The estimated background contribution from SM processes is obtained
from simulation.  After all selection requirements are applied,
\ttbar decays matching the signal topology are the dominant
background source, with W+jets events also contributing, but at a much smaller level.

Simulated events are corrected to account for effects of the trigger selection and differences between data and simulation in lepton and b-quark jet
identification efficiency. The correction factors for leptons are obtained using a high-purity data sample of
$\cPZ\rightarrow{\ell^{+}\ell^{-}}$ decays, where $\ell$ is an
electron or muon.
To account for the difference between data and simulation in the performance of the b-tagging algorithm,
we follow the method described in Ref.~\cite{btag},
which entails adding or removing a b~tag on each jet in simulated events,
based upon $\pt$- and $\eta$-dependent correction factors~\cite{btag}.

Table~\ref{tab:MCDatacomp} provides the cross sections used for each of the SM backgrounds, which, multiplied by
the integrated luminosity of 5.0\fbinv and selection efficiencies,
give the expected number of background events.
For \ttbar, we use the measured cross section from Ref.~\cite{CMSttxsec2011}.
The single top quark next-to-next-to-leading-order (NNLO) cross section is obtained from Refs.~\cite{Kidonakis:2011wy, Kidonakis:2010tc, Kidonakis:2010ux}.
The W+jets and Z+jets cross sections are computed to NNLO using the
Fully Exclusive W, Z Production (FEWZ) pQCD generator~\cite{Gavin:2010az}.
Finally, the cross section for multijets is obtained, at LO,
from \PYTHIA~\cite{ref:pythia}.

The uncertainties quoted in Table~\ref{tab:MCDatacomp} reflect statistical and systematic
sources,
which are elaborated in Section~\ref{CS:systematics}.
Although large, the estimated background is comparable to the predicted signal for a $\PWpr$ with a mass of 600\GeVcc and with $g_{L}=0$ and $g_{R}=2$~\cite{Strassler, CKYTC:2011},
given in the next-to-last row in Table~\ref{tab:MCDatacomp}.
The numbers of observed events in data in the e+jets
and $\mu$+jets channels
are presented in the last row in Table~\ref{tab:MCDatacomp}.  For $\PWpr$ signal events, the total
selection efficiency is roughly 2\%, while for the main background, \ttbar, it is roughly 0.2\%.

\begin{table*}[htbp]
\begin{center}
\topcaption{Number of events in the e+jets and $\mu$+jets channels,
for simulated background, a possible $\PWpr$ signal with a 600\GeVcc mass, and data, corresponding to 5.0\fbinv integrated luminosity. The uncertainties
include both statistical and systematic contributions from all sources 
discussed in Section~\ref{CS:systematics}.}
\begin{tabular}{|l | c | c c |}
\hline
\hline
Sample								& Cross section &	\multicolumn{ 2}{c|}{Number of events}  			\\
  & [pb] & e+jets & $\mu$+jets \\ \hline
\ttbar										& 154 $\pm$ 17~\cite{CMSttxsec2011} & 734 $\pm$ 204 & 888 $\pm$ 276		\\ \hline
Single top quark					& 85		& 32 $\pm$ 16		& 40 $\pm$ 20			\\ \hline
W+jets										& 31314	& 64 $\pm$ 32		& 49 $\pm$ 25			\\ \hline
Z+jets										& 3048 	& 8 $\pm$ 4			& 12 $\pm$ 6			\\ \hline
Multijets									& $6.7\times 10^{6}$ & 5 $\pm$ 5			& - 							\\ \hline
Total background 					&  			& 843 $\pm$ 209 & 989 $\pm$ 279 	\\ \hline \hline
Signal ($M_{\PWpr}$ = 600\GeVcc, $g_{L}=0, g_{R}=2$)& 18.2 	& 723 $\pm$ 140	& 858 $\pm$ 120		\\ \hline \hline
Data											&  			& 726 					& 904 						\\ \hline  \hline

\end{tabular}

\label{tab:MCDatacomp}
\end{center}
\end{table*}

\section{Systematic Uncertainties}
\label{CS:systematics}
The following sources of systematic uncertainty
are evaluated for their effect on
simulated signal and background yields:
jet energy corrections, b-tagging corrections,
lepton charge misidentification, lepton trigger and identification efficiencies,
and the integrated luminosity measurement.
In addition, the parameters used
in generating simulated events have associated systematic
uncertainties, as does the
procedure to make the distribution of additional interaction vertices
from pileup match simulation to data.

Uncertainties from corrections on the jet energy scale
are taken into account by shifting
the energy of each jet by $\pm$1 standard deviation~\cite{JES_official}.
The \ETm is corrected simultaneously since it is almost fully correlated
with the jet energies.
The effect on yields is about 10\% for both signal and background.
Similarly, with simulated events, varying the b-tagging corrections by their estimated uncertainty leads to a 2\% change in the event yields.

A wrong charge assignment for a reconstructed lepton could also cause its
momentum determination to be incorrect.  As a check of the significance
of this effect, study of the charge-misidentification rate with a sample of simulated
\ttbar events shows the effect to be negligible.
To match the trigger and lepton identification efficiencies in simulated samples and data, correction factors are applied.
The uncertainty on these corrections is approximately 2\% of event yields in
signal and background.

The uncertainty in the luminosity determination is 2.2\%~\cite{LumiRef2011} and affects the overall scaling of signal and background samples.

The uncertainty in the choice of PDFs used for generating simulated events is evaluated
following the PDF4LHC recommendation~\cite{pdfsys}.
We calculate an uncertainty of around 13\% in the number of expected events,
for both signal and \ttbar background.
We also evaluate the uncertainty associated with the choice of
renormalization and factorization scales and
the initial- and final-state radiation used in simulation.
Varying the $q^{2}$ scale by a factor of 4 or 0.25 results in a
7\% change in the estimated signal yield. For
\ttbar, the yield changes by about 20\%.  This systematic uncertainty is the
dominant one for the background prediction, and it reflects uncertainties in
modeling events with five or more jets, as our selection requires.

All simulated background samples are scaled using their corresponding cross sections.
For \ttbar we use the CMS measured value of $154~\pm~17$\unit{pb}~\cite{CMSttxsec2011}.
We conservatively treat the systematic uncertainty on the \ttbar 
cross section as uncorrelated with the systematic uncertainties on the \ttbar acceptance
because the cross-section
measurement employs techniques and simulated samples significantly different from those we use.
The other backgrounds are small, and their uncertainties do not make a
significant contribution to the total background uncertainty, so
conservative estimates of their uncertainties are used.
A 50\% uncertainty in the event yield is assumed for the electroweak
backgrounds (W, Z) as well as for single top quark production,
and a 100\% uncertainty is assigned to the multijet background~\cite{CMSttxsec2011}.

Simulated samples are produced with a generic distribution describing additional interactions in the same bunch crossing at high
instantaneous luminosities (event pileup). A reweighting procedure is
applied to match the pileup
conditions in the data. The uncertainty in the
reweighting procedure has an effect of about 1\% on the event
yields for both simulated signal and background.

\section{Limits on \texorpdfstring{\PWpr}{W'} Production}
\label{sec:Limits}

We perform a counting experiment where the number of observed events
($N_\text{obs}$) is compared with the number of expected events from background ($N_\text{exp}$).
From Table~\ref{tab:MCDatacomp}, we have $N_\text{obs} = 726$ and $N_\text{exp}
= 843 \pm 209$ for the e+jets channel, and $N_\text{obs} = 904$ and $N_\text{exp}
= 989 \pm 279$ for the $\mu$+jets channel.  We observe no significant
excess above the SM background expectation in the two channels.

From a comparison of the observed and estimated numbers of events, we calculate an upper limit on the cross section of $\PWpr$ production as a function of mass.
A 95\% confidence level (CL) upper limit is calculated using the
CL$_S$
technique~\cite{Read:2002hq,Junk:1999kv}.
Any theoretical uncertainty in the LO $\PWpr$ cross section is not included.
Systematic uncertainties in the luminosity, jet energy scale,
b-tagging efficiency, pileup, and lepton ID efficiency are
taken to be 100\% correlated between signal and background.
Systematic uncertainties due to PDFs and factorization scales are taken
to be uncorrelated for signal and the leading \ttbar background.
All systematic uncertainties are assumed to follow log-normal
distributions.

\begin{figure}[htbp]
\begin{center}
\includegraphics[width=0.5 \textwidth]{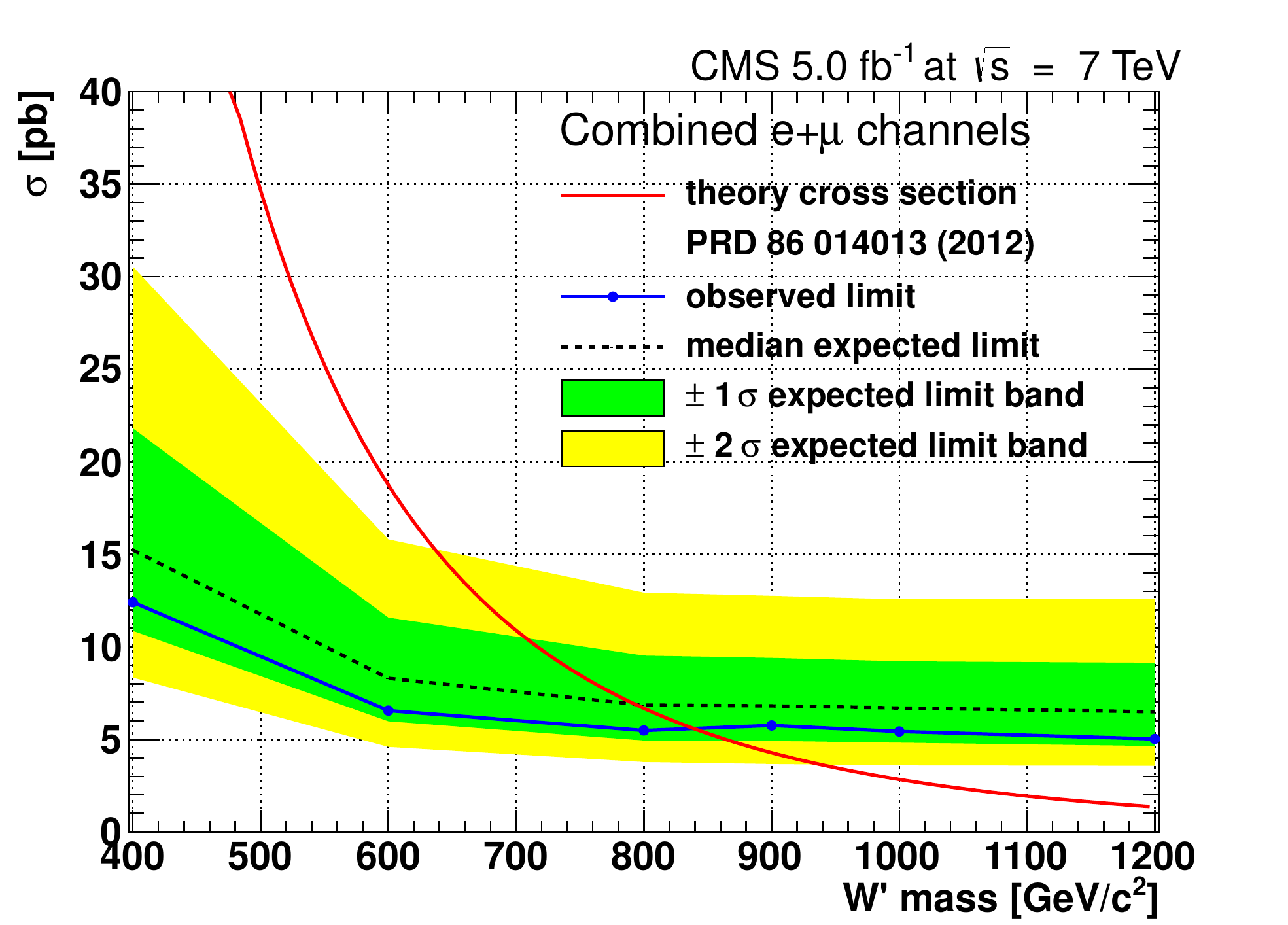}
\caption{The 95\%~CL expected and observed limits on LO $\PWpr$ production
for $g_{L}=0$ and $g_{R}=2$
as a function of the $\PWpr$ boson mass for e+jets and $\mu$+jets channels combined.}
\label{fig:kk_combination}
\end{center}
\end{figure}

The 95\%~CL upper limit on the number of selected signal events is 581 for the
combination of the e+jets and $\mu$+jets channels with $\PWpr$ coupling constant values $g_{L}=0$ and $g_{R} = 2$. Figure~\ref{fig:kk_combination} shows
the corresponding 95\%~CL upper limit on the $\PWpr$ production. For this $\PWpr$ model, we exclude $\PWpr$ masses below 840\GeVcc.

To provide comparison between two important theoretical benchmark points of the $\PWpr$ model, we also calculate limits on the cross section for $\PWpr$ masses of
600 and 800\GeVcc and $g_{R}$ values of 2 and $\sqrt{2}$, as
shown in Table~\ref{diffCouplingW}.

\begin{table}[htbp]
\begin{center}
\topcaption{$\PWpr$ cross-section limits for $g_{R} = 2$ and
$\sqrt{2}$ (with $g_{L}=0$) for the combined e+jets and $\mu$+jets channels.
The \PWpr\ model includes both s- and t-channel production. Thus, the
acceptance changes with the coupling, and the cross section does not simply scale with the coupling.}
\begin{tabular}{|l|c c||c c|}
\hline
\hline
$\PWpr$ mass & \multicolumn{ 2}{c||}{600\GeVcc} & \multicolumn{ 2}{c|}{800\GeVcc}
\\ \hline
Coupling  $g_{R}$ & 2 & $\sqrt{2}$ & 2 & $\sqrt{2}$ \\ \hline
Cross section [pb] & 18.2 & 6.3 & 6.5 & 2.1 \\ \hline
Acceptance (comb.) & 1.8\% & 2.2\% & 2.1\% & 2.8\% \\ \hline
Expected limit [pb] & 8.3 & 6.3 & 6.9 & 5.3 \\ \hline
Observed limit [pb] & 6.6 & 5.0  & 5.5 & 4.3 \\ \hline
\hline
\end{tabular}
\label{diffCouplingW}
\end{center}
\end{table}

\section{Search for \texorpdfstring{\PWpr}{W'} Asymmetry}
\label{Sec:CS_strategy}
As an independent cross check, we attempt the reconstruction of the
$\PWpr$ mass and derive an asymmetry to check whether there is any
indication of signal in the data, as suggested by Ref.~\cite{Strassler}.
We exploit the charge asymmetry in $\PWpr$ production, a key feature of this theoretical model.

To reconstruct the $\PWpr$, we first reconstruct two top quarks.
Three jets in the event are used to reconstruct one top quark,
with the jets being considered to match the decay chain $\cPqt\rightarrow\PW + \cPqb$,
$\PW\rightarrow j + j$, where $j$ is a jet resulting from the hadronic decay of the
W boson.
Out of the many three-jet combinations
possible in each event, we choose the one in which a pair of jets gives an invariant mass closest
to the $\PW$ boson mass~\cite{pdg}
and the three jets give an invariant mass closest to the top quark mass~\cite{pdg}. From the
lepton, one jet, and \ETm, we reconstruct the second top quark, following the
decay chain $\cPqt\rightarrow\PW + \cPqb$, $\PW\rightarrow \ell + \nu$.
Again, out of the several combinations with the jets in each event,
we choose the one giving an invariant mass closest to the top quark mass.
Though the event selection requires at least one b-tagged jet, b tagging is not considered when choosing
the jets used for top reconstruction.
Next, the highest-\pt, non-b-tagged jet not used in the top reconstruction is
labeled as the candidate d-quark jet.

\begin{figure}[htbp]
\begin{center}
\includegraphics[width=0.45 \textwidth]{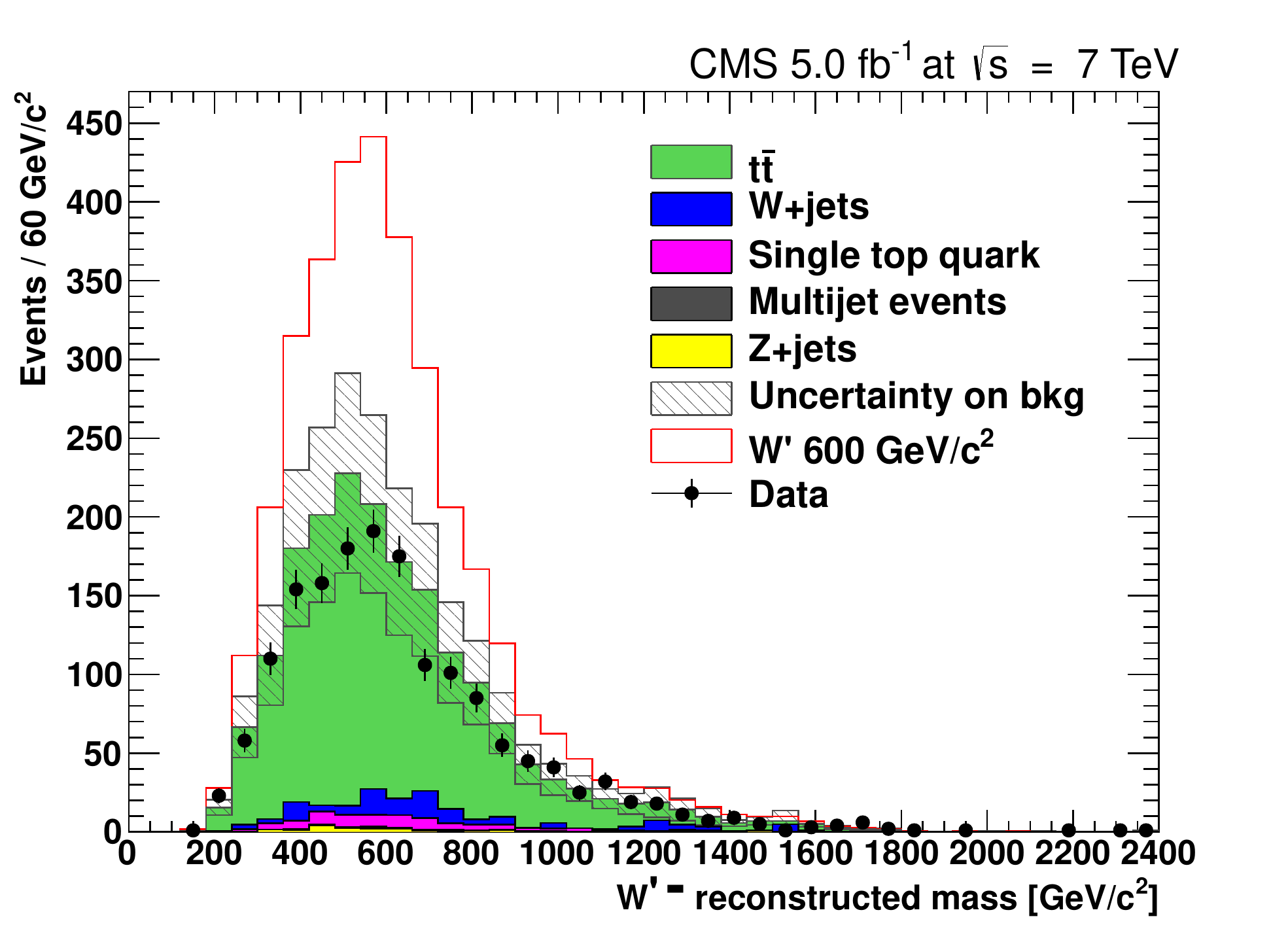}
\includegraphics[width=0.45 \textwidth]{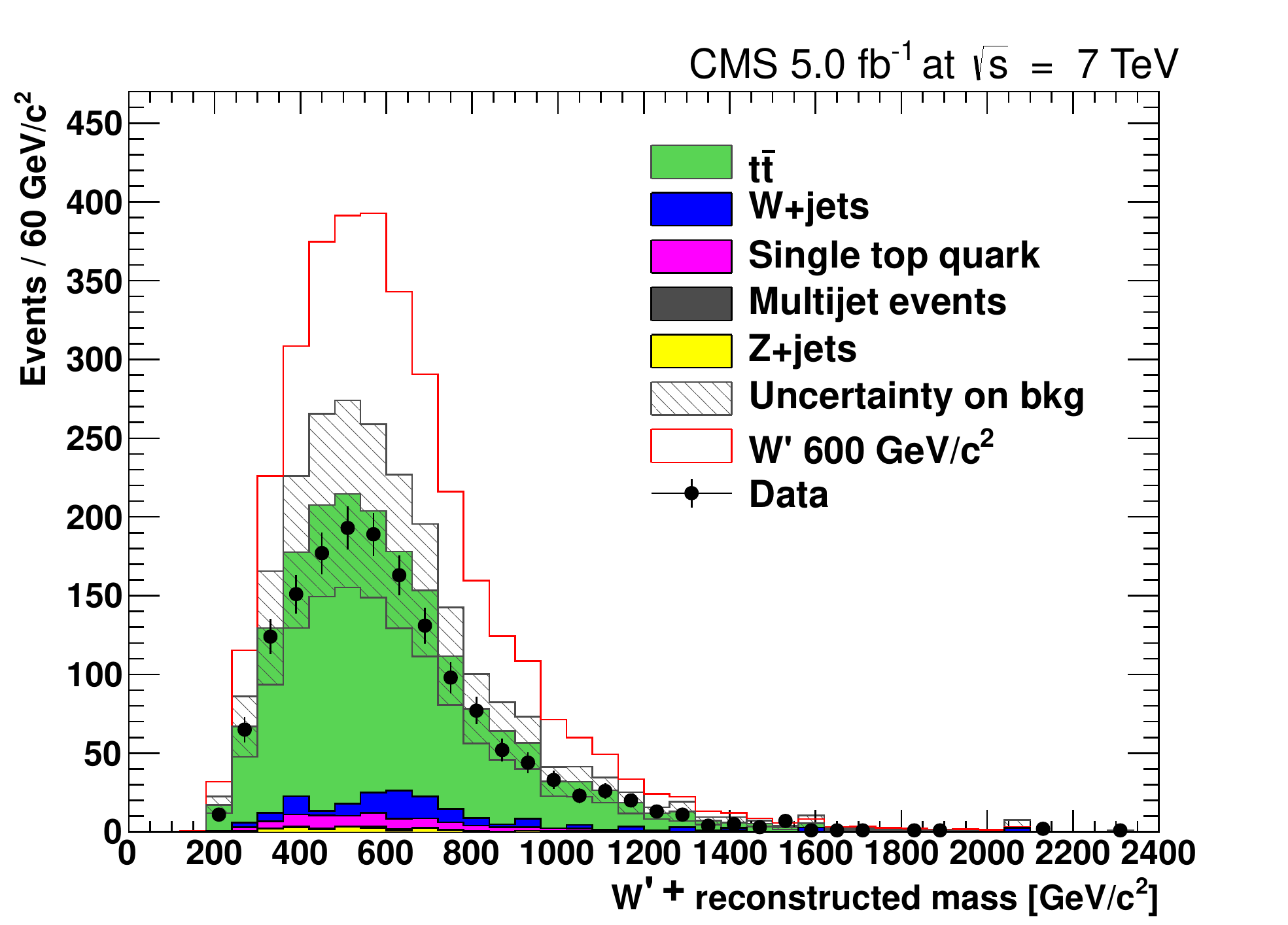}
\caption{Invariant mass distributions for (\cmsLeft) $\pwprm$ candidates
and (\cmsRight) $\pwprp$ candidates. The figures show a comparison between the background prediction, with candidates reconstructed from simulated signal events stacked on top, and data. Uncertainty bands represent statistical and systematic uncertainties on the background prediction.}
\label{fig:wpmass}
\end{center}
\end{figure}

Each of the two top candidates are combined in turn with the
d-quark jet so that two $\PWpr$ candidates are reconstructed for each selected event.
The charge of the $\PWpr$ candidate that has the lepton in its decay chain is determined by the charge
of the lepton, while the other $\PWpr$ candidate has the opposite charge.
In a true $\PWpr$ event, only the $\PWpr$ candidate that matches the true $\PWpr$ charge
could possibly be correctly reconstructed, and thus,
because of the charge asymmetry in $\PWpr$ production, the
$\pwprm$ candidate  is more likely to be correctly reconstructed than the
$\pwprp$ candidate.
Figure \ref{fig:wpmass} shows the invariant mass distributions for both
$\PWpr$ candidates reconstructed from every selected event in
data and the simulated background. Included with both distributions is the expected signal for a $\PWpr$ boson with a mass of 600\GeVcc.
The invariant mass distribution for the $\pwprm$ candidates shows a high peak
around the $\PWpr$ mass, while the distribution for the $\pwprp$ candidates has a lower, more rounded peak. The simulated $\PWpr$ events provide this shape
difference, since, for these events, $\pwprm$ candidates cluster more
around the $\PWpr$ mass while $\pwprp$ candidates, likely to be mis-reconstructed, create
a broader distribution.
A window of 200\GeVcc width around the $\PWpr$ mass contains, for
$\PWpr$ signal with a mass of 600\GeVcc,
about 42\% of 
$\pwprm$ candidates compared with only about 34\% of $\pwprp$ candidates.
The background components of both distributions look identical since the
additional jet originates from initial- or final-state radiation
and there is no preference in combining it with either top candidate.
The data distributions agree within the uncertainties with the background model.

\begin{figure}[htbp]
\begin{center}
\includegraphics[width=\cmsFigWidthB]{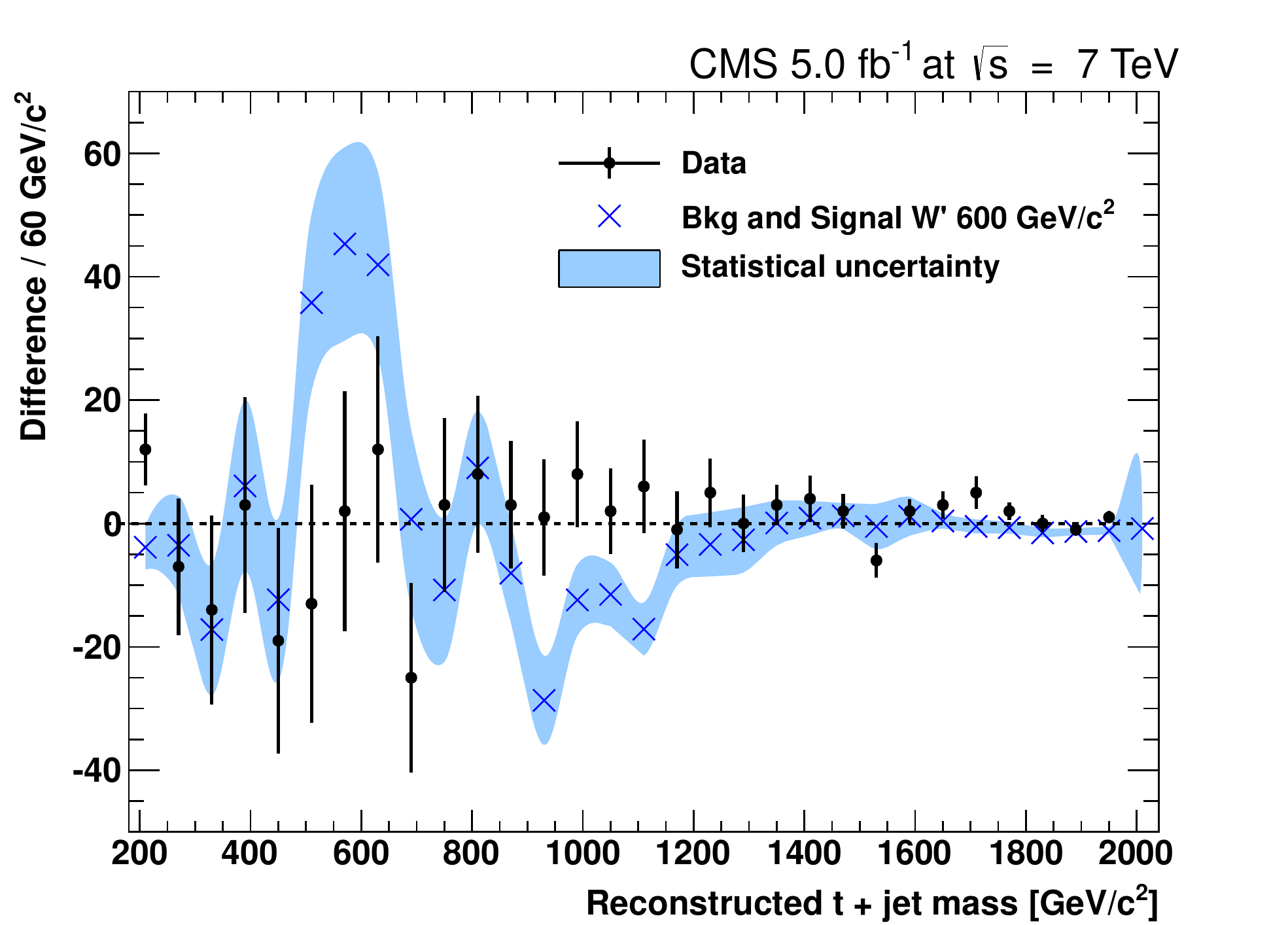}

\caption{Difference in yields for $\pwprm$ and
$\pwprp$ candidate invariant mass distributions.
Data are compared with the combination of
background and a simulated $\PWpr$ signal with a mass of 600\GeVcc
and $g_{L}=0$ and $g_{R}=2$.  The shaded blue band indicates
the statistical uncertainty of the signal and background combination.}
\label{fig:diff_mcdat}
\end{center}
\end{figure}

To illustrate the asymmetry of the $\PWpr$ model, we calculate the difference in yields for
the $\pwprm$ and $\pwprp$ invariant mass distributions.
The result is shown in Fig.~\ref{fig:diff_mcdat}.
The data are represented by black points, and blue X's show the expected difference for a combination of
background and a simulated $\PWpr$ signal at a mass of 600\GeVcc.
The shaded blue band represents the statistical uncertainty for the
combined signal and background.
The predicted
signature of $\PWpr$ events, as seen in Fig.~\ref{fig:diff_mcdat} is
a bump at the $\PWpr$ mass and a dip at higher mass, due to the fact that the $\pwprm$
mass distribution is more tightly concentrated within a narrow peak compared with the broader
distribution seen for $\pwprp$.
In contrast, the data are statistically consistent with a flat distribution.

\section{Summary}\label{sec:limitSummary}

A search has been performed by the CMS collaboration for a $\PWpr$ boson via the process
$\cPqd + \Pg \to \cPqt + \PWpr$, $\PWpr\to \cPaqt+\cPqd$.
This model represents one possible explanation for the \ttbar forward-backward asymmetry seen at the Tevatron.
The data showed no significant deviation from the standard model prediction.
A counting experiment set
a 95\%~CL limit on the $\PWpr$ production cross
section as a function of mass.
This $\PWpr$ model with $g_{L}=0$ and $g_{R}=2$ has been excluded below a mass of 840\GeVcc in the combined \Pe+jets and $\Pgm$+jets channels.  In addition, no statistically significant indication of the predicted $\PWpr$ mass distribution asymmetry has been observed in the data.

During the final stages of publication of this work, a related article has
appeared~\cite{Endo:2012mi},
suggesting that interference effects were not properly taken into account in the
theoretical model used in our analysis, with a possible result being the alteration of
the limits we quote. The interference effects discussed in
Ref.~\cite{Endo:2012mi} arise mainly in
diagrams with t-channel $\PWpr$ exchange, but these effects do not contribute
significantly in the region of phase space chosen by our full selection.

\section*{Acknowledgments}
We thank Matt Strassler, Simon Knapen, and Yue Zhao for the simulation and computation of the cross sections at 7 TeV for the $\PWpr$ model used in this analysis.
We congratulate our colleagues in the CERN accelerator departments for the excellent performance of the LHC machine. We thank the technical and administrative staff at CERN and other CMS institutes, and acknowledge support from: FMSR (Austria); FNRS and FWO (Belgium); CNPq, CAPES, FAPERJ, and FAPESP (Brazil); MES (Bulgaria); CERN; CAS, MoST, and NSFC (China); COLCIENCIAS (Colombia); MSES (Croatia); RPF (Cyprus); MoER, SF0690030s09 and ERDF (Estonia); Academy of Finland, MEC, and HIP (Finland); CEA and CNRS/IN2P3 (France); BMBF, DFG, and HGF (Germany); GSRT (Greece); OTKA and NKTH (Hungary); DAE and DST (India); IPM (Iran); SFI (Ireland); INFN (Italy); NRF and WCU (Korea); LAS (Lithuania); CINVESTAV, CONACYT, SEP, and UASLP-FAI (Mexico); MSI (New Zealand); PAEC (Pakistan); MSHE and NSC (Poland); FCT (Portugal); JINR (Armenia, Belarus, Georgia, Ukraine, Uzbekistan); MON, RosAtom, RAS and RFBR (Russia); MSTD (Serbia); SEIDI and CPAN (Spain); Swiss Funding Agencies (Switzerland); NSC (Taipei); TUBITAK and TAEK (Turkey); STFC (United Kingdom); DOE and NSF (USA).

Individuals have received support from the Marie-Curie programme and the European Research Council (European Union); the Leventis Foundation; the A. P. Sloan Foundation; the Alexander von Humboldt Foundation; the Belgian Federal Science Policy Office; the Fonds pour la Formation \`a la Recherche dans l'Industrie et dans l'Agriculture (FRIA-Belgium); the Agentschap voor Innovatie door Wetenschap en Technologie (IWT-Belgium); the Council of Science and Industrial Research, India; the Compagnia di San Paolo (Torino); and the HOMING PLUS programme of Foundation for Polish Science, cofinanced from European Union, Regional Development Fund.

\bibliography{auto_generated}   

\cleardoublepage \appendix\section{The CMS Collaboration \label{app:collab}}\begin{sloppypar}\hyphenpenalty=5000\widowpenalty=500\clubpenalty=5000\textbf{Yerevan Physics Institute,  Yerevan,  Armenia}\\*[0pt]
S.~Chatrchyan, V.~Khachatryan, A.M.~Sirunyan, A.~Tumasyan
\vskip\cmsinstskip
\textbf{Institut f\"{u}r Hochenergiephysik der OeAW,  Wien,  Austria}\\*[0pt]
W.~Adam, T.~Bergauer, M.~Dragicevic, J.~Er\"{o}, C.~Fabjan\cmsAuthorMark{1}, M.~Friedl, R.~Fr\"{u}hwirth\cmsAuthorMark{1}, V.M.~Ghete, J.~Hammer, N.~H\"{o}rmann, J.~Hrubec, M.~Jeitler\cmsAuthorMark{1}, W.~Kiesenhofer, V.~Kn\"{u}nz, M.~Krammer\cmsAuthorMark{1}, D.~Liko, I.~Mikulec, M.~Pernicka$^{\textrm{\dag}}$, B.~Rahbaran, C.~Rohringer, H.~Rohringer, R.~Sch\"{o}fbeck, J.~Strauss, A.~Taurok, P.~Wagner, W.~Waltenberger, G.~Walzel, E.~Widl, C.-E.~Wulz\cmsAuthorMark{1}
% ...
\vskip\cmsinstskip
\dag:~Deceased\\
1:~~Also at Vienna University of Technology, Vienna, Austria\\
% ...

\end{sloppypar}
\end{document}